\documentclass[structabstract]{aa}
\usepackage{graphicx}
\usepackage{txfonts}
\usepackage{natbib}
\usepackage{textcomp}
\bibpunct{(}{)}{;}{a}{}{,}

\begin{document}

\def\C10{C10}

\title{The bright end of the z $\sim$ 7 UV Luminosity Function from a wide and deep HAWK-I survey}

\author{ M. Castellano \inst{1}
  \and
  A. Fontana \inst{1}
  \and
  D. Paris \inst{1}
  \and
  A. Grazian \inst{1}
  \and
  L. Pentericci \inst{1}
  \and
  K. Boutsia \inst{1}
  \and
  P. Santini \inst{1}
  \and
  V. Testa \inst{1}
  \and
  M. Dickinson  \inst{2}
  \and
  M. Giavalisco  \inst{3}
  \and
  R. Bouwens  \inst{4}
  \and
  J.-G. Cuby\inst{5}
  \and
  F. Mannucci  \inst{6}
  \and
  B. Cl\'ement\inst{5} 
  \and
  S. Cristiani  \inst{7}
  \and
  F. Fiore \inst{1}
  \and
  S. Gallozzi \inst{1}
   \and
  E. Giallongo \inst{1}
  \and
  R. Maiolino  \inst{1}
  \and
  N. Menci \inst{1}
  \and
  A. Moorwood \inst{8}
  \and
  M. Nonino \inst{7}
  \and
  A. Renzini \inst{9}
  \and
  P. Rosati \inst{8}
  \and
  S. Salimbeni \inst{3}
  \and
  E. Vanzella  \inst{7}  
   }

\institute{INAF - Osservatorio Astronomico di Roma, Via Frascati 33,
00040 Monteporzio (RM), Italy \and NOAO, 950 N. Cherry Avenue, Tucson, AZ 85719, USA \and Department of Astronomy, University of Massachusetts, 710 North Pleasant Street, Amherst, MA 01003 \and 
Lick Observatory, University of California, Santa Cruz, CA 95064, USA,  \and Laboratoire d'Astrophysique de Marseille, OAMP, Universit\'e Aix-Marseille \& CNRS, 38 rue Fr\'ed\'eric Joliot Curie, 13388 Marseille cedex 13, France \and INAF - Osservatorio Astrofisico di Arcetri, Largo E. Fermi 5, I-50125 Firenze, Italy \and INAF - Osservatorio Astronomico di Trieste, Via G.B.
Tiepolo 11, 34131 Trieste, Italy  \and European Southern Observatory, Karl-Schwarzschild-Str. 2, D-85748 Garching, Germany \and INAF - Osservatorio Astronomico di Padova, Vicolo dell'Osservatorio 5, I-35122 Padova, Italy  }

   \offprints{M. Castellano, \email{castellano@oa-roma.inaf.it}}

   \date{Received .... ; accepted ....}

\titlerunning{The bright end of the z $\sim$ 7 UV Luminosity Function}
\authorrunning{M. Castellano et al.}

\abstract
%aims heading (mandatory)
{}{We perform a deep search for galaxies in the redshift range
  $6.5\le z\le 7.5$, to measure the evolution of the number density of
  luminous galaxies in this redshift range and derive useful
  constraints on the evolution of their luminosity function.  }
% methods heading (mandatory)
{We present here the second half of an ESO Large Programme, which
  exploits the unique combination of area and sensitivity provided in
  the near--IR by the camera Hawk-I at the VLT.  We have obtained  $\sim30$ observing hours with Hawk-I in the $Y$-band of two
  high galactic latitude fields. We combined the $Y$-band data with deep $J$ and $K$ Hawk-I observations, and with FORS1/FORS2 $U$, $B$, $V$, $R$, $I$, and $Z$ observations to select $z$-drop galaxies having  $Z-Y>1$, no optical detection and flat $Y-J$ and $Y-K$ colour terms.}
% results heading (mandatory)
{We detect 8 high-quality candidates in the magnitude range
  $Y=25.5-26.5$ that we add to the $z$-drop candidates selected in two Hawk-I pointings over the GOODS-South field. We use this full sample of 15 objects found in $\sim 161~ arcmin^{2}$ of our survey to constrain the average physical properties and the evolution of the number density of $z\sim7$ LBGs. A stacking analysis yields a best-fit SED with photometric redshift  $z=6.85^{+0.20}_{-0.15}$   and an $E(B-V)=0.05^{+0.15}_{-0.05}$. We compute a binned estimate of the $z\sim7$ LF and explore the effects of photometric scatter and model uncertainties on the statistical constraints.
 After accounting for the expected incompleteness through MonteCarlo simulations, we strengthen our previous finding that
  a Schechter luminosity function constant
  from z=6 to z=7 is ruled out at a $\gtrsim$99\% confidence level, even including the effects of cosmic variance. For
  galaxies brighter than $M_{1500}=-19.0$, we derive a luminosity density
  $\rho_{UV}= 1.5^{+2.1}_{-0.8}\times 10^{25} erg ~ s^{-1} ~ Hz^{-1} ~
  Mpc^{-3} $, implying a decrease by a factor 3.5 from $z=6$ to
  $z\simeq 6.8$. We find that, under standard assumptions, the emission rate of ionizing photons coming from UV bright galaxies is lower by at least a factor of two than the value required for reionization. Finally, we exploit deep Hawk-I $J$ and $K$ band observations to derive an upper limit on the number density of  $M_{1500} \lesssim -22.0$ LBGs at $z\sim8$ ($Y$-dropouts).
}  
{}


\maketitle
%
%________________________________________________________________

\section{Introduction}

The search and study of galaxy populations at very high redshift is one of the
most promising research areas of today astrophysics and
cosmology. It derives its importance on two different and interrelated
aspects: 1) The estimate of the UV photon
budget provided by star-forming galaxies and its role on the reionization of
the universe at $z>6$; 2) The study of the formation and the physical properties of the first
bulding blocks of present-day galaxies.

There is observational evidence that the Universe is highly ionized at $z\sim6$ \cite[e.g.][]{Fan2006,Totani2006}, in agreement with 
the latest WMAP estimates of the Thomson optical depth \citep{Komatsu2010}, although significant uncertainties remains on the homogeneity \cite[e.g.][]{Mesinger2009} and on the exact timeline of the reionization process \cite[e.g.][]{Gallerani2006}. Whether the UV light emitted by star-forming galaxies is capable of reionizing the Universe by these epochs remains an open question that should be answered through the analysis of large samples of high redshift objects.

The search for high-redshift star forming galaxies has been carried out so far
mainly with renditions of the Lyman Break, or ``drop-out'' technique that has been proved to be extremely efficient
at redshift from 2 to 6 \cite[e.g.][]{Steidel1995,Steidel1999,Adelberger2004,Dickinson2004,Giavalisco2004,Ouchi2004,Bouwens2007,Mclure2009}, or through narrow-band studies
targeting the Ly$\alpha$ emission \cite[e.g.][]{Iye2006,Kashikawa2006,Ouchi2009b}. The application of the  Lyman Break technique  at $z>6$ has been performed, at first, in small
areas with deep near-IR $J+H$ NICMOS data \cite[e.g.][]{Bouwens2004}, and it has recently acquired
momentum thanks to the installation of the WFC3 camera onboard of the
Hubble Space Telescope yielding to a sample of tens of faint Lyman Break galaxies (LBGs)  \citep{Bouwens2009b,Oesch2009b,Mclure2009b,Bunker2009,Yan2009,Wilkins2009,Wilkins2010}. 
In the meantime, ground based surveys \cite[][\C10 hereafter]{Ouchi2009,Capak2009,Hickey2009,Castellano2010}, along with refined analysis of archival NICMOS observations \citep{Bouwens2010} have expanded the number of bright LBGs known.

The basic feature of the high redshift galaxy population that can be analysed 
through the present datasets is its  UV
luminosity function (LF).
The current picture of the evolution of the UV LF points to a
factor of 6-11 decrease in the number density of UV bright galaxies from $z\sim 3$ to $z\sim 6$
\cite[e.g.][]{Stanway2003,Shimasaku2005,Bouwens2006a}, although some
uncertainties are still present
on the exact amount of evolution in the different parameters of the Schechter
function \citep{Dickinson2004,Giavalisco2004,Sawicki2006,Iwata2007,Yoshida2006,Bouwens2006a,Bouwens2007,Beckwith2006}. 
At redshift above 6, most of the analysis indicates a strong evolution in the
LF, mainly through a dimming of the characteristic magnitude
$M_{*}$  and/or a decrease of the normalization factor $\phi$ \citep{Bouwens2008,Mclure2009b,Ouchi2009,Yan2009,Castellano2010,Bouwens2010}. The recent WFC3-based analysis by \citet{Oesch2009b} also found evidence for a steep faint-end
($\alpha \sim -1.8$), in agreement with the predictions of theoretical models \citep{Trenti2010,Salvaterra2010}. LBGs searches around lensing clusters have also been performed,
 finding discrepant results
that highlight the many challenges and uncertainties in these investigations \citep{Richard2006,Richard2008,Bradley2008,Bouwens2009,Zheng2009}.
Along with improved constraints on the LF at $z\sim7$, the latest analysis of the
WFC3 data have also provided a first estimate of the evolution at $z\sim8-9$
that points to a further decrease in the LBG number density, and thus in the
total amount of UV photons produced by young stars at these early epochs.

The discrepancies among different works, both at $z\sim3-5$ and at $z\sim6-9$ are most probably due to the effect
of cosmic variance \cite[e.g.][]{Trenti2008,Robertson2010}, but also to the difficulties in avoiding systematic effects in the different
estimates of completeness level, contamination from lower redshift interlopers,  volume elements, and redshift
distributions in the various samples \citep{Stanway2008b}, all worsened by the known
degeneracy among the parameters adopted to fit the LF.

The strong decrease observed in the
UV emission coming from relatively bright sources seems to imply that reionization cannot be
 explained on the basis of UV bright galaxies only.
An increased number of low luminosity galaxies indicated by the steep faint end of the Schechter LF might play a decisive role in
the reionization process. Large and reliable samples of high-z galaxies both at the bright and at the faint end of the LF are thus necessary to shed light on this issue, and, possibly to highlight the need to search for even more intriguing sources of the reionizing radiation with future facilities \cite[see e.g.][]{Venkatesan2003,Madau2004}.

Latest surveys have also given the opportunity of analysing the physical properties of high redshift galaxies, whose knowledge is also a decisive factor to understand the very role of these sources in the reionization process. Recent studies have given the first estimates of masses, ages and SFRs for single $z \gtrsim 6.5$ objects, the first constraints on the stellar mass density at these epochs, and have also raised an interesting debate on the possibilty that the first galaxes might be characterized by peculiar properties, like a very low dust content, nearly primordial metallicity or top-heavy stellar initial mass functions \citep{Finkelstein2009,Bouwens2010,Gonzalez2010,Labbe2010,Salvaterra2010,Schaerer2010}.

To give an answer to some of the above problems we are using the new VLT IR imager Hawk-I
\citep{Pirard2004,Casali2006,Kissler2008}, to conduct a deep, medium area survey in the $Y$ band over four independent pointings, aimed at the detection of relatively bright LBGs at $6.5<z<7.5$.
Thanks to the extreme efficiency and large field of view (7.5$\times$7.5 arcmin) of Hawk-I, it is possible to easily reach $Y\sim26.5$ AB at $>5\sigma$ (roughly corresponding to $M_{1500}=-20.5$ at $z=7$), over large areas in a reasonable amount of time (15 hrs).

In \C10 we discussed the results of the first half of our survey,
covering a large fraction of the GOODS-S field, and we 
 estimated a statistically significant ($\sim$99 \% c.l.) decrease, with respect to $z\sim6$, of the number density of UV-bright galaxies.
In this paper we present the $z\sim7$ candidates found in the second half of the survey,
covering two other independent fields. We will constrain the evolution of the LF combining this new sample with the GOODS one. 

The paper is organised
as follows. In Section 2 we present the imaging set and the 
multiwavelength catalogue; in Section 3 the LBG 
selection criteria and the potential interlopers affecting the selection are discussed; 
in Section 4 we present our final sample of candidate $z$-drop LBGs. In  Section 5 we discuss a stacking analysis of all the $z$-drop galaxies found in the four Hawk-I pointings, that are used to
constrain the $z>6$ UV LF in Section 6. In Section 7 we derive an upper limit on the number density of very bright $z\sim8$ LBGs. 
A summary of our methods and results is provided in Section 8.

Throughout the whole paper, observed and rest--frame magnitudes are in
the AB system, and we adopt the $\Lambda$-CDM concordance model
($H_0=70km/s/Mpc$, $\Omega_M=0.3$, and $\Omega_{\Lambda}=0.7$).

\section{Data}
\label{dataset}
\begin{figure}
   \centering

   \includegraphics[width=9.0cm]{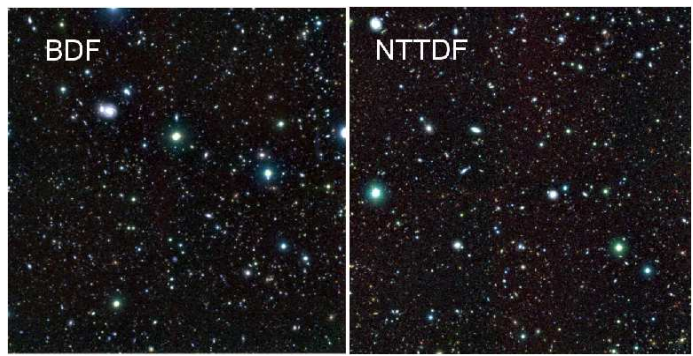}

   \caption{Colour-composite image of the BDF (left) and NTTDF (right) fields, created using the
weighted mean of Hawk-I $Y$, $J$ and $K$ images as red, the FORS2 $Z$ as green, and the weighted mean of FORS1/FORS2 optical images as blue.}
         \label{fields}
\end{figure}
\begin{table*}
\centering
\caption{BDF - Observations}
\label{tabBDF}

\begin{tabular}{cccccc}
\hline
Filter & Instr. & Exp. Time (s)& Seeing (arcsec) & Mag. Limit$^a$  \\
\hline
V-High& FORS2 & 13800& 0.75 & 29.1 \\
R-Special& FORS2 & 11600 & 0.63 &   29.3\\
I-Bessel& FORS2 & 4800 & 0.70 &   27.8 \\
Z-Gunn& FORS2 & 64800 & 0.59 &  28.6 \\
Y-Open& HAWK-I & 56940 & 0.52 & 28.3 $^b$\\
J-Open& HAWK-I & 18720 & 0.54 &  26.5 \\
Ks-Open& HAWK-I & 30060 & 0.44 &  26.0  \\

\hline
\end{tabular}
\\
\smallskip
\begin{tabular}{l}
a - S/N=1\\
b - Y=26.5 at S/N=5 \\
\end{tabular}
\\

\end{table*}

\begin{table*}
\centering
\caption{NTTDF - Observations}
\label{tabNTTDF}

\begin{tabular}{cccccc}
\hline
Filter & Instr. & Exp. Time (s) & Seeing (arcsec) &Mag. Limit$^a$  \\
\hline
U-Bessel& FORS1 & 32876& 0.84& 27.8 \\
B-Bessel& FORS1 & 16064 & 0.56&  28.9 \\
V-Bessel& FORS1 & 10500 & 0.47&  29.0 \\
R-Special& FORS2 &  14000& 0.79 &  28.4 \\
I-Bessel& FORS2 & 7830 & 0.61&  28.0 \\
Z-Gunn& FORS2 & 46386 & 0.60 &   28.4  \\
Y-Open& HAWK-I & 54180 & 0.49 &  28.3$^b$ \\
J-Open& HAWK-I & 14400 & 0.47 &  26.7 \\
Ks-Open& HAWK-I & 24720 & 0.39&  26.3 \\

\hline
\end{tabular}
\\
\smallskip
\begin{tabular}{l}
a - S/N=1\\
b - Y=26.5 at S/N=5 \\
\end{tabular}
\\

\end{table*}
\subsection{Observations}

This work is based on deep $Y$--band images obtained with the IR camera Hawk-I
at the VLT, and on deep optical FORS2 observations. We use data collected through a dedicated ESO Large Programme in 2008 and 2009. 
The first set of data, covering two adjacent regions
of the GOODS-S field has been presented in \C10. Here we present the analysis
of two other pointings (Fig.~\ref{fields}), chosen for the wealth of deep, public observations previously exploited by other authors to search for $z\sim4-6$ LBGs: 
the BDF field at Ra=336.98\textdegree, Dec=-35.17\textdegree \citep{Lehnert2003}, and the
New Technology Telescope Deep Field (NTTDF) at Ra=181.36\textdegree, Dec=-7.72\textdegree \citep{Arnouts1999,Fontana2000,Fontana2003}.
The total exposure time is 15h49m for BDF and 15h03m
for the NTTDF in the $Y$ band.

The $Y$ band images were reduced  using standard techniques
for IR data - flat fielding, sky subtraction among consecutive frames,
and final coaddition. The
reduction procedure, which is described in detail in our first paper \C10, has
been specifically designed to enhance the reliability of the images at the
faintest fluxes, and to get rid of persistence effects and cross-talk resonances.

We  determine an FWHM of 0.52 $\pm 0.01$ arcsec ($\simeq$ 4.9 pixels) in the final
coadded BDF image and 0.49 $\pm 0.01$ arcsec ($\simeq$ 4.6 pixels) in the NTTDF one.
Image zeropoints were computed using the standard stars observed during the same
night and at similar airmasses. Reference fluxes were converted to the photometric system 
and filter set used in this paper, as described in \C10.

We  obtained the absolute r.m.s. maps for each pointing  by computing the r.m.s. in each individual image (using the
Poisson statistics and the instrumental gain) and propagating
self-consistently this r.m.s. over the whole data reduction process.
The typical $5\sigma$ magnitude in one arcsec$^2$ is in the range
26.7-26.8 over more than 60\% of the whole image, and $>26.2$ in 85\%
of the image - the rest of the images being shallower because of the
gaps between the four Hawk-I chips.

A wide wavelength coverage  is needed to reliably
select high redshift LBGs excluding lower redshift interlopers and red and
dusty galaxies at intermediate redshift. To this aim we obtained or re-reduced
deep observations of both fields ranging from the blue to the near-IR, matching
all images  to the $Y$ band pixel-size and astrometric solution. 

Along with the main $Y$--band pointings, we also acquired deep $J$ and $Ks$ Hawk-I observations of both fields. 
We also obtained $\sim 7$ hours of FORS2 $Z$ band coverage for each field, that we coadded
with the already existing FORS2 images \citep{Fontana2003} to reach the required depth.
We also re-reduced the archive $U$, $B$, $V$, $R$, $I$ FORS2 and FORS1 observations of the NTTDF, and the FORS2 $R$ and $I$ images of the
BDF. Finally, we obtained $\sim 4$ hours of $V$-FORS2 observations on the BDF. The full dataset is presented in  Tab. \ref{tabBDF} and
Tab. \ref{tabNTTDF}.

\subsection{The photometric catalogue}
\label{catalogue}

\subsubsection{Detection}
\label{detection}
We obtained the photometric catalogue using the SExtractor code V2.5
\citep{Bertin1996} and the $Y$ band as detection image with the
r.m.s. map derived as described above. Since high redshift galaxies are
almost unresolved in ground-based images, and SExtractor's \verb|MAG_BEST| are known to
underestimate the total flux of faint objects ($Y > 24$ in our case), we chose to use aperture-corrected
total magnitudes.
We computed aperture magnitudes in a 2 FWHM diameter 
and corrected them to total magnitudes adopting aperture corrections from bright non-saturated stars in
each field. While this choice might give slightly underestimated fluxes for
the more extended high redshift candidates, we can easily take into account
this systematic through the
simulations that we use to estimate the LF (Sect.~ \ref{Montecarlo}) that are
based on the observed profile of LBGs
 with known spectroscopic redshifts $5.5<z<6.2$ \citep{Vanzella2009} in the GOODS-S
ACS images.

 We  optimised the SExtractor parameters involved in the detection process through the analysis of a
`negative' image as discussed in \C10, adopting the set of
parameters that minimises the ratio between 'negative' and 'positive'
detections at the faint end of the number counts.
As expected, we find that the best parameters for faint objects detection on BDF and NTTDF
$Y$-band images are the same adopted for the similar set of images over GOODS-South:
we require 10 contiguous pixels each at $S/N>0.727$,
corresponding to a $2.3\sigma$ detection, and we restrict the analysis
to the regions where the r.m.s is less than $\sim$ 1.5 times the lowest
value.  With this choice of parameters, we do not find any detection on the negative
images at $Y<26.2$, and a fraction of negative detections less than 5\% of the real ones at fainter magnitudes. However, \textit{a posteriori}, the latter value overestimates the actual rate of spurious detections. Indeed,
  all spurious sources should appear as ``drop-out'' candidates with a
  single-band detection. On the contrary, all the $Y>26.2$ objects in our $z$-drop sample are confirmed by detections in other IR bands. Indeed, as we discuss also in \C10, the test on the negative image is probably influenced by non-trivial issues concerning the subtraction of the background or a potential asymmetry in the noise
  distribution.

\subsubsection{The multicolour catalogue}\label{catal}
A multiwavelength catalogue containing self-consistent
magnitudes in all available bands was built running SExtractor in dual mode using the
$Y$-band Hawk-I image as the detection image with the detection parameters
indicated above.
 Aperture fluxes were computed within a 2FWHM aperture
 and converted to total applying appropriate aperture corrections in each band.  

The typical $1\sigma$ limiting magnitudes in a 2FWHM aperture are in the
range $27.8-29.3$ for the optical bands,  $J \sim 26.5-26.7$, and  $Ks \sim 26.0-26.3$.
The corresponding $1\sigma$ limiting magnitude in the $Z$ band, which is used to define the 'dropout' selection, is 
$\sim 28.6$ in BDF and $\sim 28.4$ in the NTTDF (see Tab. \ref{tabBDF} and
Tab. \ref{tabNTTDF}). 

For each field we defined the total areas where the image depth is sufficiently homogeneous. 
The candidates found in this area will be used for the evaluation of the LF.
We used $Y$-band detected objects only in the regions selected
on the basis of the negative image test explained above. In addition, we also masked borders, CCD defects and noisiest
regions in the other images of our data-set. 
The areas selected in this way correspond to $\sim 71$\% of the  $Y$-band coverage in the
BDF, and $\sim 56$\% in the NTTDF (due to strong vignetting in the $Z$-band image). As a result, the total area used for $z$-drop detection amounts to ~$71.7$~ arcmin$^2$.  We will subtract to this value the fraction of area covered by lower redshift objects ($\sim$9\%) to estimate effective volumes in Sect.~\ref{LF} and Sect.~\ref{z8}.

\section{The selection of z$>6.5$ galaxies}\label{Selection}
\subsection{The dropout criterion}\label{drop}

We select candidate $z>6.5$ galaxies using the ``drop-out'' technique adapted
to our filter set and imaging depth. 

\begin{figure}
   \centering
  \includegraphics[width=9cm]{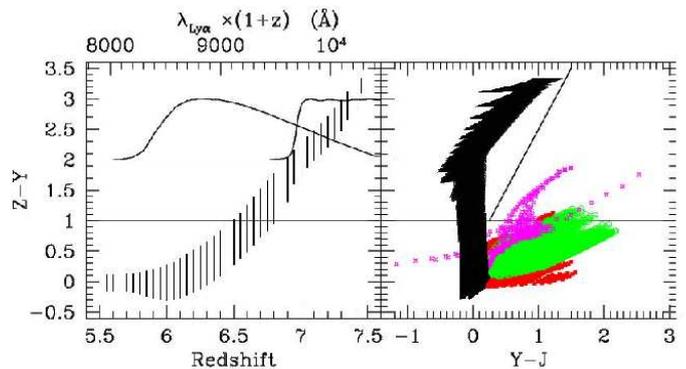}
   \caption{{\it Left}: $Z-Y$ colour of star forming galaxies as a function of redshift. In the
upper part, the efficiency curve of the two filters is shown, computed
at observed wavelength of a Lyman-$\alpha$ emission at the corresponding
redshift. {\it Right}: $Z-Y$ vs. $Y-J$ colour diagram showing the expected colours of LBGs (same as in left panel, black points), passively evolving galaxies (red squares) and
     reddened starbursts (green circles) at $1.5<z<4$ and cool dwarf stars from the templates of \citet{Tsuji2004} (magenta stars). 
     Galaxy colours are computed according to CB07 models, see text for details on the adopted parameters. In both panels lines indicate the relevant colour selection criteria discussed in Sect. \ref{Selection}
   }
         \label{diagram}
\end{figure}

In order to individuate the appropriate selection criteria we estimated the expected colours of high-redshift star forming galaxies (black points in the right panel of Fig.~ \ref{diagram}) on
the basis of the models of Charlot
and Bruzual 2007 \citep{Bruzual2007a,Bruzual2007b} with the same range of free
parameters as in \C10: Metallicity: 0.02, 0.2 and 1 $Z_\odot$; age from 0.01 Gyr
to the maximal age of the Universe at a given $z$; E(B-V)=0...0.2
\citep{Calzetti2000}. Ly-$\alpha$ rest-frame
equivalent width in the range 0-200 \AA.
Intergalactic absorption following \citet{Madau1995}. The same range of model
parameters will be used as baseline for the MonteCarlo simulations used
to estimate the LF in a self consistent way, see Sect. \ref{LF}.

As shown in the left panel of Fig.~ \ref{diagram}, galaxies at
$z>6.5$ show an increasing  $Z-Y$ colour which is due to the sampling within these two
filters of the  sharp drop
shortward of the Lyman-$\alpha$, where most of the photons are absorbed
by the intervening HI in the intergalactic medium. The drop in the flux
observed shortward of the $Y$ band is analogous to the one used to select star
forming galaxies at lower redshifts, like $i$-drops at $z\sim5$, $V$-drops at $z\sim4$
etc.: the major difference with respect to the standard Lyman break technique being that
the $Y$ band does not sample the
continuum around $1500$ \AA ~ but a region shortward of it,
contaminated by both the larger IGM absorption at $z>6$ and by the Lyman-$\alpha$ emission
line.  
These effects can only be accurately accounted for by realistic imaging
simulations, as we discuss in detail in \C10 and in  
section ~\ref{Montecarlo} of this paper.
Following this test, we choose $Z-Y>1$ as our main selection criterion to
select  $z>6.5$ galaxies. Given that the $Z$-band observations, as well
as the optical ones, used in the present paper are slightly shallower than
the GOODS-ACS ones, we limit our selection at $Y<26.5$ instead of the $Y<26.8$
adopted in \C10. 

The selection of $z$-drop galaxies cannot be solely based on the $Z-Y$
colour, since other classes of objects can
display a red $Z-Y$ colour similar to that of $z>6.5$ galaxies. 
Selection criteria, both in the optical and in the IR bands, are thus necessary to individuate a reliable sample of $z\sim7$ galaxies.

\subsection{IR colour selection}\label{ircolours}
We tailored our IR colour selection to exclude any possible
contamination in our  $z$-drop sample from known classes of lower redshift objects:

 i) We modelled passively evolving galaxies and dusty starburst galaxies at
$z>1.5$  with a suitable set of spectral synthesis models. We
use the same CB07 library as for high redshift galaxies to predict the
colours of such objects at $1.5<z<4$, using a combination of short
star formation exponential timescales ($0.1-1$ Gyrs) and ages $>1$ Gyr
to reproduce passively evolving galaxies, and constant star-forming
models with $0.5<E(B-V)<1.5$ \cite[adopting a][extinction law]{Calzetti2000} 
for the dusty starbursts. As shown in Fig. \ref{diagram} (right panel) these
galaxies also show a large IR colour term.

To exclude these objects, we 
adopt  the same additional criteria on IR colours as in \C10 :~
(Z-Y) $>$ (Y-K); ~(Z-Y) $>$ 0.5+2.0(Y-J); ~(Y-J)$<$ 1.5;~(Y-K)$<$ 2.0.

 ii) Cool ($T_{eff}<1500$ K), low-mass stars, and substellar objects of the T
spectral class have infrared spectra that are
dominated by the  $CH_{4}$ and $H_2O$ absorption bands and
by $H_{2}$ resonant absorption
\cite[e.g.][]{Chabrier2005,Burgasser2006} that produce a sharp break in
their IR colours. We used the most up-to-date estimate of the T-dwarfs
number density (as observed in the $J$ band) of Burgasser et al 2007 to compute the expected number of faint, cool
dwarfs in our fields. Adopting an average $Y-J$ colour of 0.8 mags estimated from
the catalogue of observed dwarfs compiled by \citet{Leggett2010}, and considering the dependence on galactic
latitude as in \citet{Burgasser2004}, we estimated that $\sim$ 0.6 stars of spectral types T0-T8 with $Y<26.5$ are
expected to fall in each of our fields. However the exact number of expected contaminating dwarfs
depends on the still uncertain parameters constraining the IMF and their
spatial distribution inside the disk and the halo
of the Galaxy: a pessisimistic estimate (see \C10) gives a nearly double surface density
of cool dwarfs in our pointings.
For this reason we used the synthetic spectral libraries by \citet{Tsuji2004} \cite[see also][]{Tsuji2002,Tsuji2005}
to check whether it is possible to define selection criteria discriminating between LBGs and cool dwarfs.
As shown in the right panel of Fig. ~\ref{diagram}, cool dwarfs appear redder than $z$-dropouts in the $Y-J$ colour, and
the $Y-J$ criterion we adopt allows us to exclude these objects from our selection window. 
We note that the brown dwarf discovered in the NTTDF by \citet{Cuby1999},
having $Z-Y\sim2$ and no optical detection, is consistently
excluded from our high-z sample on the basis of its  $Y-J$ colour.

 iii) Finally, we cross-checked each object selected according to the above criteria against
variability, by analysing images acquired at different epochs. The BDF observations have been split in two separated epochs 
with a 3 months gap (September and December 2009), while the NTTDF have been observed during four runs in January, February, April and May 2009.
We verified that all the objects in our sample are clearly detected, and that they have a consistent total flux (within $2\sigma$), in the different epochs. In the NTTDF case, we checked that a detection $> 3\sigma$ of the faintest candidate ($Y\sim26.5$) was possible in the two epochs with larger integration time.

We summarise here the full set of colour selection criteria:
\begin{eqnarray*}
Y &<& 26.5 \\ (Z-Y) &>& 1.0\\
(Z-Y) &>& (Y-K)\\ (Z-Y) &>& 0.5+2.0(Y-J)\\ 
(Y-J)&<& 1.5\\ (Y-K)&<& 2.0\\
\end{eqnarray*}

\subsection{Comparison with the GOODS-ACS dataset}

In our analysis of the GOODS-South field we exploited the ACS V2.0 $B$, $V$, $I$, $Z$ observations (M. Giavalisco and the GOODS Team, in preparation) to select $z$-drop galaxies and to exclude lower redshift interlopers showing significant detection in the optical bands. The main concern we have to consider to provide a $z$-drop selection as clean as the one in the GOODS field regards the difference in resolution between FORS2 optical observations of BDF and NTTDF and their corresponding ACS-GOODS images we used to remove interlopers from the colour-selected sample.

Indeed, in \C10 we found
that a sample of galaxies selected with IR 
criteria only is populated also by faint contaminants showing 
significant detection in filters covering
wavelengths shorter than the redshifted Lyman limit at $z>6$ ($U$, $B$, $V$, $R$, $I$) where high redshift LBGs are not expected to present any flux. 
 These objects are, in most cases, clearly extended, but their spectral energy
 distributions  cannot be reproduced by a straightforward application of the
 CB07 models. While determining their nature is beyond the scope of the present analysis,
we note that they might be faint galaxies with a very blue
continuum whose SED is altered by strong emission
lines such as in unobscured AGNs, or in star-forming galaxies like the
blue compact dwarf galaxies \citep{Izotov2004,Izotov2007} or the ultra
strong emission line galaxies \cite[USELs,][]{Hu2009}. Potential contamination of z$\sim 7$ samples due to an unknown class of objects with no optical detection ($<2\sigma$) has also been suggested by \citet{Capak2009}. Their objects are brighter than those found in our fields but display similar colours. Given the unknown nature of this contaminants, at present, the only feasible approach is to adopt more stringent criteria on the optical non-detections. Follow-up spectroscopy of z$\sim 7$ candidates and faint contaminants is needed to fully evaluate the impact of this population on high redshift studies.

In our analysis of the GOODS-South field we adopted
 very strict selection criteria in the blue bands in order to exclude these
contaminants,  measuring the S/N ratios in
small apertures (0.6'') exploiting the high resolution of ACS images. 

In order to obtain optical selection criteria as effective as the ones used with the GOODS dataset, we
performed tests computing S/N ratios and
photometry on the GOODS-ACS images degraded and smoothed to the depth/seeing
of the BDF and NTTDF corresponding ones.
We then re-selected GOODS dropouts on ``mock'' BDF/GOODS and NTTDF/GOODS
catalogues built in the same way as the real BDF and NTTDF catalogues. We verified that the criteria already adopted in the GOODS fields are effective in the NTTDF case: $S/N<2\sigma_{S/N}$ in all the optical bands and $<1\sigma_{S/N}$ in at least
four of them.
In the BDF, given the absence of $U$ and $B$ images and the slightly shallower $I$ imaging, we adopted the conservative criterion $S/N<1\sigma_{S/N}$ in all the
 optical bands. We verified that this criterion allows us to safely remove all those objects, up to $Y$=26.5, that have been
 verified to be lower redshift contaminants on the basis of GOODS-ACS and,
 whenever possible, UDF-ACS photometry. 

The parameter $\sigma_{S/N}$ indicated above is the r.m.s. of the S/N
distribution estimated, as in \C10, dropping random apertures in portions of the images
free of detected objects. This procedure allows us to take into account the sky noise
distribution and the presence of faint, undetected foreground objects at the
same time.

\section{Detected z$>6.5$ galaxies}

\begin{figure*}[!ht]
   \centering
   \includegraphics[width=16cm]{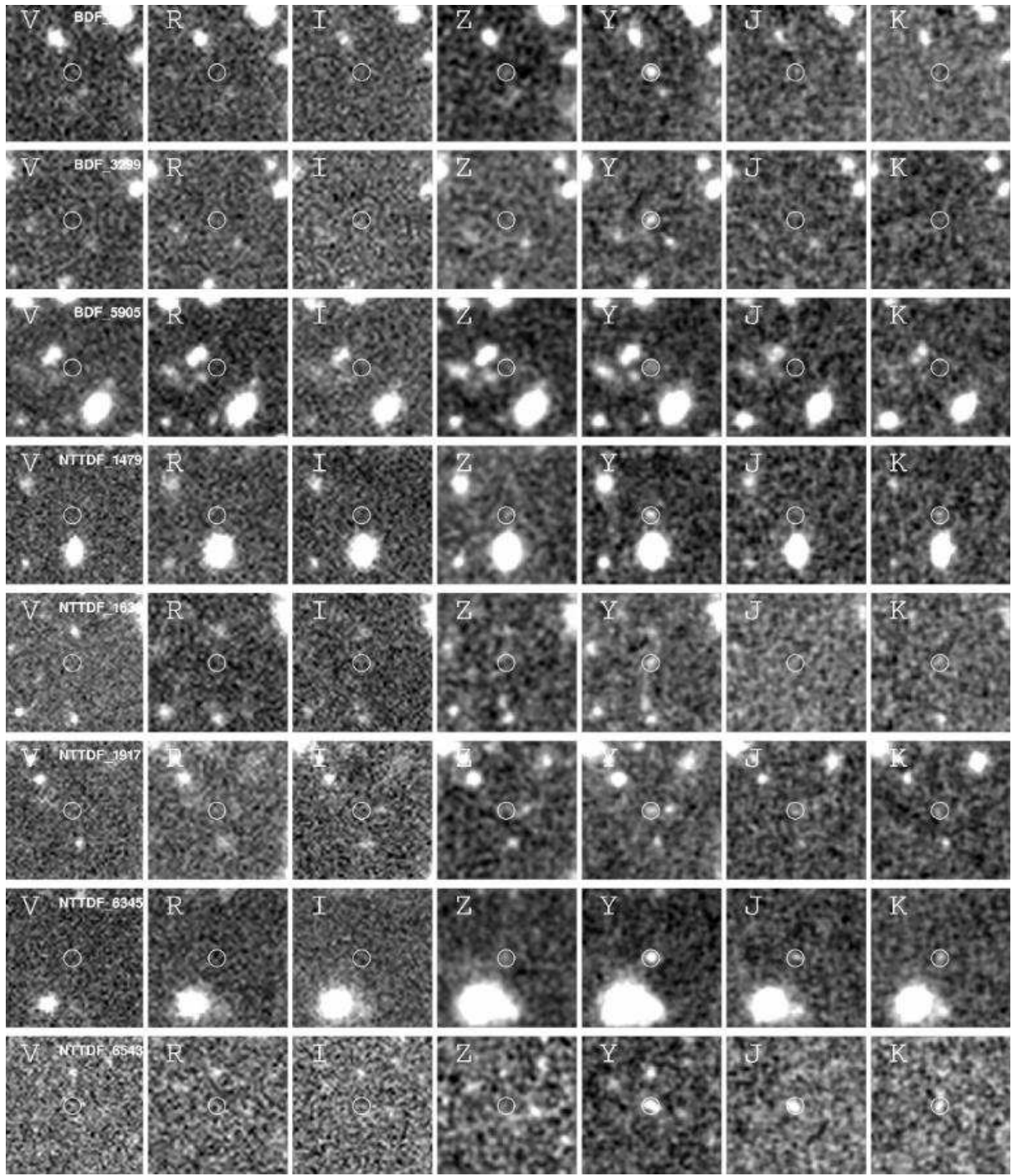}
   \caption{Thumbnails showing the images of the 8 selected high-redshift 
candidates in the different observed bands.}
         \label{thumb}
\end{figure*}

Adopting the selection criteria outlined above we find a total of eight candidates, three in the BDF and five in the NTTDF field,
whose coordinates, $Y$ magnitudes and $Z-Y$ colours are listed in Tab~\ref{data}. Thumbnails of the candidates are presented in Fig.~\ref{thumb}.
We note that two of them are clearly detected, and two others are marginally detected (S/N$\sim2$), in the $Z$ band.
Three of the five candidates present in the NTTDF are also detected at S/N$\sim2-4$ in the $J$ and $K$ bands, thanks to the slightly deeper images available for this field (see Tab~\ref{tabNTTDF}). We also verified that each candidate is undetected in the image obtained as the weighted sum of its $V$, $R$ and $I$ observations. 

As a final check we performed a stacking of all the objects in the available images.
This test allows us to confirm the non-detection in the optical images, and to obtain a clear 
detections in the $J$ (S/N$\sim5$) and $K$ band (S/N$\sim4$) stacked images. The stacked object shows an average colour
$Z-Y\simeq 1.6$.

In the following sections we will combine this sample $z$-drop candidates found in the BDF and NTTDF pointings with the sample discussed in \C10, obtained from the two pointings over the GOODS-South field, to find their average properties through a stacking analysis, and to constrain their LF. The GOODS $z$-drop sample includes seven candidates in the range $Y\sim25.5-26.7$, selected through colour selection criteria analogous to the ones outlined above, whose reliability has also been checked on the available IRAC and NICMOS observations. 

\begin{figure}[!ht]
   \centering
   \includegraphics[width=8.6cm]{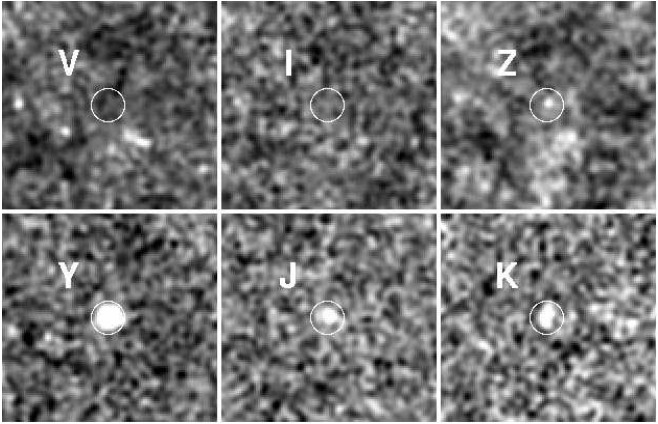}

   \caption{Thumbnails showing the stacked images of the 15
     high-redshift candidates selected in the GOODS1, GOODS2, BDF and NTTDF fields. The observed bands are shown in
     the legends.}
         \label{thumb_stack}
\end{figure}

\begin{figure*}[!ht]
   \centering
   \includegraphics[width=14cm]{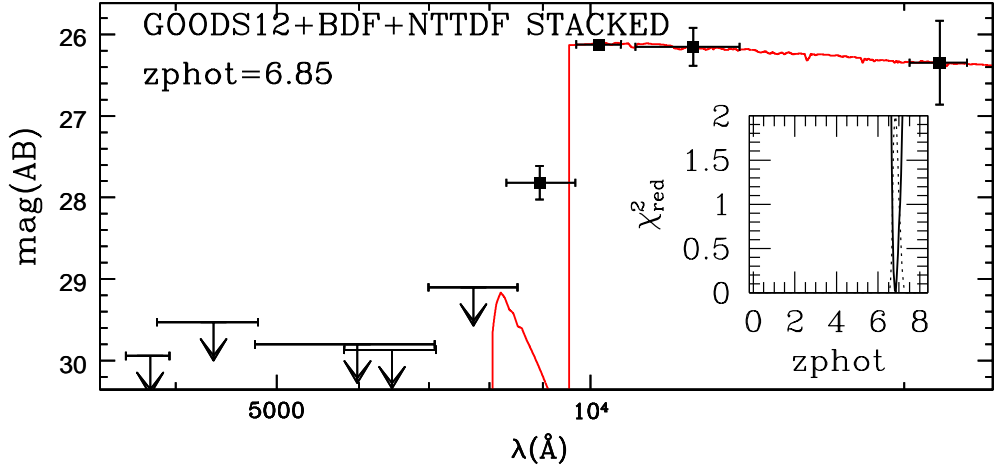}
   \caption{Best-fit SED to the stacked photometry, with relevant
     photometric redshift at $z=6.85$.}
         \label{sed_stack}
\end{figure*}

\begin{table}
\caption{Candidates in the BDF and NTTDF fields}
\label{data}
\centering
\begin{tabular}{cccccc}
\hline
ID & R.A. (deg) & DEC. (deg)& Y & Z-Y & S/N (Y)\\
\hline
BDF\_521& 336.9444 & -35.1188& 25.86& 2.13& 10.2\\
BDF\_3299& 337.0511 &-35.1665 & 26.15& $>$2.4& 7.8\\
BDF\_5905& 337.0230 & -35.2094& 26.24&1.20 & 7.6\\
NTTDF\_1479& 181.3429 & -7.6813 & 26.12 & 1.97& 8.4\\
NTTDF\_1632& 181.3857 &  -7.6835 & 26.47 & 1.61& 6.1 \\
NTTDF\_1917& 181.3212 & -7.6877 & 26.32 & 1.58 & 7.1\\
NTTDF\_6345& 181.4039 & -7.7561 & 25.46 & 1.45& 15.6\\
NTTDF\_6543& 181.3834& -7.7595 & 25.75 & $>$2.6& 12.0\\
\hline
\end{tabular}
\end{table}

\section{Mean properties of Hawk-I $z\sim7$ galaxies}\label{stacked}

We perform a weighted mean of the images in the available filters from the $U$ to the $K$ band for all the 15 objects detected in the Hawk-I fields. We did not attempt a similar stacking of the IRAC images, since most of our GOODS candidates are partially or extremely blended with other foreground sources, and the candidates in the other fields are either not covered by IRAC observations, or they are present in shallower exposures with respect to GOODS. We matched the ACS images to the Hawk-I PSF and masked all the foreground objects surrounding the candidates in each image.
The stacked object shows an $S/N \gtrsim 5$ detection in the $Z$, $J$ and $K$ bands and a non-detection in all the optical bands, corresponding to an ($optical - Y$) colour of $>4$ magnitudes.  We use the magnitudes estimated for the stacked object to find the photometric redshift and physical parameters through our photo-z code \citep{Giallongo1998,Fontana2000} exploiting a $\chi^2$ minimisation procedure 
to find the best-fitting spectral template to the observed colours among the full CB07 library. While the ACS optical filters used in the GOODS field have different passbands with respect to the FORS2 ones used in BDF and NTTDF fields, this is not a significant concern since they all span a wavelength range where no flux is expected for $z>6$ objects. In turn, the small difference between FORS2 and ACS $Z$-band filters does not provide significant variations in the redshift selection window defined by the $Z-Y$ colour which is the main constraint to the photometric redshift.
The resulting SED provides a unique photometric
redshift solution at $z=6.85^{+0.20}_{-0.15}$. Relevant thumbnails and SED are shown in Figure~
\ref{thumb_stack} and ~\ref{sed_stack}. Given the absence of IRAC, most physical parameters  are largely unconstrained, apart from the E(B-V) parameter whose estimate is mostly based on the $Y-J$ and $Y-K$ colours. We find that our stacked SED is fitted by an $E(B-V)=0.05^{+0.15}_{-0.05}$ at a 68\% confidence level. This value is consistent with the E(B-V) distribution obtained from the analysis of z$\sim7-8$ objects by \citet{Finkelstein2009} and by \citet{Schaerer2010}. Our best fit E(B-V) indicates a low dust content for $z\sim7$ galaxies, in agreement also with the best-fit $A_V$ values found by \citet{Gonzalez2010} and by \citet{Labbe2010} for the mean SED of their $z$-drop samples, and with the blue UV continuum slope measured by \citet{Bouwens2010b}.

\section{The evolution of the LF}
\label{LF}

\subsection{MonteCarlo simulations}\label{Montecarlo}
When small galaxy samples are used to constrain the high-redshift LF, it is necessary to exploit detailed imaging simulations to appropriately treat the systematic effects arising from faint object detection, and from the application of colour selection criteria.
To this aim we use the CB07 synthetic libraries described in Sect.~\ref{Selection}
to produce, for each field, 
a set of  $\sim 8\times10^5$ simulated LBGs with redshift in the range $5.5<z<8.0$ and observed magnitudes computed in
the same filter set used for the observations. These
galaxies are placed at random positions of the 
$Y$-band images, and catalogs are extracted exactly as in the original
frames. To avoid an excessive crowding in the simulated images,
we include only 200 objects each
time, after masking the regions of the images where real objects are present. 
As in \C10, we randomly assign to each of our simulated galaxies the light profile of one of the four most distant spectroscopically confirmed
LBGs observed with ACS in GOODS \cite[$z=5.5-6.2$,][]{Vanzella2009}, after convolving it with the relevant Hawk-I PSFs. 

\subsection{Stepwise LF}
\label{stepwise}
The magnitude range covered by our survey, $Y\simeq25.5-26.7$ roughly corresponds to the UV continuum magnitude range at $M \lesssim M_*$. For this reason, we first perform a binned estimate of the number density of the Hawk-I $z$-drop galaxies through the stepwise method \cite[see, e.g.][]{Bouwens2008}. The stepwise estimate is a non-parametric method based on the assumption that the rest-frame LF of galaxies
  can be approximated by a binned distribution, where the number
  density $\phi_i$ in each bin is a free parameter.
To evalute also the potential systematics and the effects of observational uncertainties in this kind of estimates, we use two different procedures to compute the stepwise LF. The first one is the procedure commonly adopted in the literature based on the average  relation between the observed $Y$ and the UV continuum magnitude at 1500\AA~ ($M_{1500}$), and on an estimate of the completeness in the different UV magnitude bins. The second, more conservative, procedure takes in consideration the uncertainties in the $Y$-$M_{1500}$ conversion due to photometric scatter, to the redshift distribution and to the intrinsic properties of different galaxy models. In a separate work we will combine this stepwise analysis  with similar estimates at fainter and brighter magnitudes to determine the Schechter parameters at $z\sim7$ in a self-consistent way (Grazian et al. 2010, in preparation).

\begin{figure}[!ht]
   \centering
    \includegraphics[width=8cm]{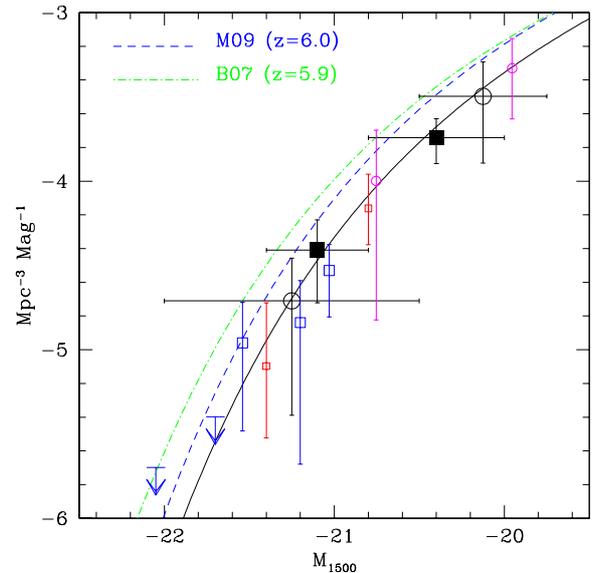}
   \caption{Number densities in two rest-frame magnitude intervals estimated
     for our Hawk-I data set in a stepwise form with a standard $Y$-UV conversion of the observed number counts as discussed in the text (black filled squares), or with a $\chi^2$ method considering also photometric and model uncertainties (black empty circles). Other points are from \citet{Bouwens2010} (NICMOS, red empty squares),
     \citet{Ouchi2009} (SUBARU, blue empty squares and upper limits) and \citet{Oesch2009b} (WFC3-UDF, magenta empty circles). 
For a comparison we show the recent determinations
     of the LF at $z\sim 6$ by \citet{Bouwens2007} (B07, green dot-dashed line) and \citet{Mclure2009} (M09, blue dashed line). The black solid line is the best-fit LF obtained by combining the stepwise points shown in the figure with new determinations of the binned densities from WFC3-UDF and WFC3-ERS data (Grazian et al. in preparation).}
         \label{Fig_stepwise}
\end{figure}

\subsubsection{Stepwise LF from the average Y-UV relation}

Through a linear regression we compute the average $Y-M_{1500}$ relation at the median redshift of our sample (z=6.8) for the CB07 models of $z$-drop galaxies. We then divide our sample in two bins centered at $M_{1500}=-21.1$ and $M_{1500}=-20.4$, and use the imaging simulations to estimate the completeness of our selection. Finally, we convert the redshift dependent completeness distribution into effective volumes of our survey at these magnitudes. 
The values of the stepwise LF estimated in this way are reported in Tab.~\ref{Tab_stepwise} and plotted as filled squares in Fig~\ref{Fig_stepwise}, with vertical error bars given by Poisson uncertainties in the number counts. The horizontal error bars indicate the relevant magnitude range of each bin.

\begin{table}
\caption{Stepwise determination of the UV LF}
\label{Tab_stepwise}
\centering
\begin{tabular}{cc}
\hline
Mag. Range & $\phi ~ (10^{-4} Mpc^{-3}~ mag^{-1})$\\
\hline
$-21.4<M_{1500}<-20.8$& $0.39 \pm 0.20$\\
$-20.8<M_{1500}<-20.0$& $1.81 \pm 0.54$\\
\hline
\end{tabular}
\end{table}

\begin{figure}[!ht]
   \centering

   \includegraphics[width=8cm]{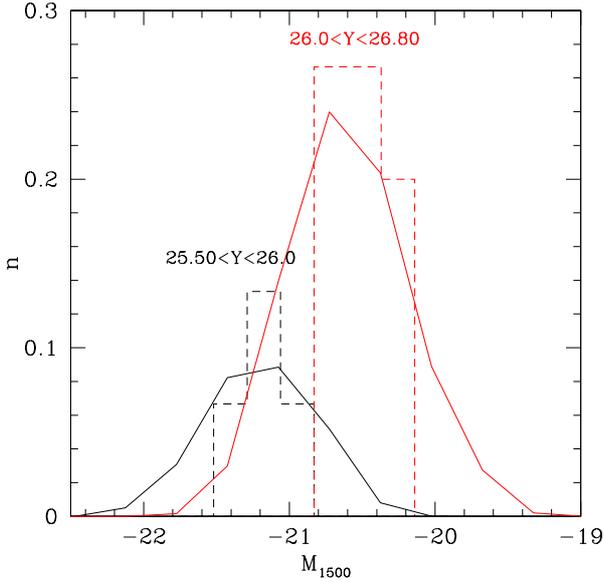}
   \caption{The normalized distribution of UV continuum magnitudes (estimated from the average $Y-M_{1500}$ relation) for the 15 Hawk-I candidates divided in two bins (dashed histograms). The solid curves show the expected distributions of UV magnitudes for objects in the same observed ranges when photometric uncertainties are taken into account through MonteCarlo simulations.}
         \label{Fig_scatter}
\end{figure}

\subsubsection{Introducing photometric uncertainties in the Stepwise LF}
A more conservative estimate can be computed assuming a stepwise LF made of three bins in the wider magnitude range $-22.0<M_{1500}<-19.0$. This interval takes into account the photometric scatter and the variation of the $Y-M_{1500}$ relation with redshift and galaxy models (see Fig.\ref{Fig_scatter}).  We assume a fixed, constant, reference density $\phi_{ref}$, and we exploit the set of simulations described in 
  Sect.~\ref{Montecarlo} to compute for each field the distribution of observed magnitudes
  originating in each rest-frame bin for LBGs in the redshift range sampled by our colour selection. The simulated number counts are then scaled to the relevant observed areas and summed together. Finally, we find the combination of binned densities $\phi{_i}=w_i \cdot \phi_{ref}$, that best reproduces
  the total number counts of our survey, where $w_i$ are multiplicative factors to the reference density that we determine by comparing observed and simulated distributions through a simple $\chi^2$ test. We plot as black empty circles in Fig~\ref{Fig_stepwise} the two bins at $M_{1500}<-19.8$. The third, faintest, bin at $M_{1500}>-19.8$ yields only a conservative upper limit and it is not represented in the figure, but it is anyway necessary in this procedure to consider the effect of Malmquist bias. Vertical error bars indicate the statistical uncertainties given by the $\chi^2$ test.
\smallskip

 The two methods give consistent results, and they are in perfect agreement with other stepwise estimates in the same magnitude range (see Fig.~\ref{Fig_stepwise}). However, the error bars and the relevant magnitude range are much larger when using the $\chi^2$ minimization procedure. While an average conversion from observed to rest-frame magnitudes, along with an estimate of effective volumes, can provide a first order-of-magnitude estimate of the binned number density of LBGs, we emphasize that significant statistical uncertainties can arise due to photometric scatter, and to the different relation between $Y$ and UV continuum magnitudes for different galaxy models and redshifts.

\subsection{Maximum Likelihood LF}\label{ML}

\begin{figure}
   \centering
 \includegraphics[width=9cm]{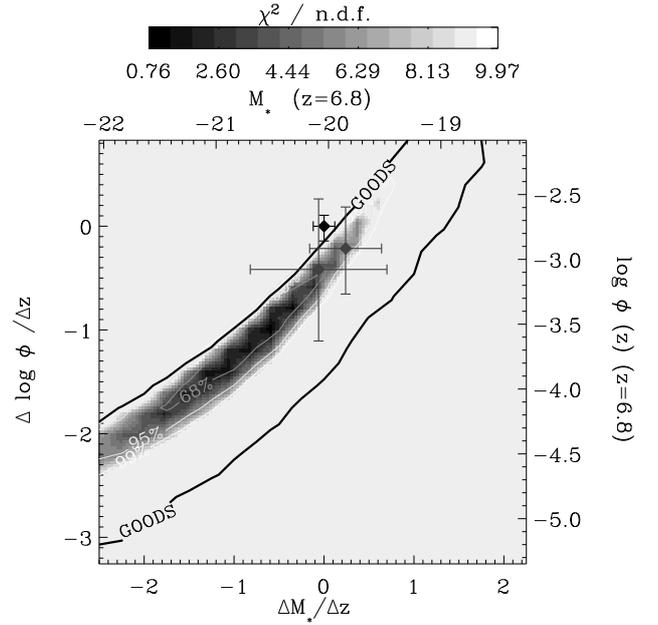}
   \caption{$\chi^2$ contour levels for the $ dlog(\phi)/dz$, $ M_*'$
     parameters derived for the Schechter--like LF considering all the four Hawk-I fields. The lower and left axis
     refer to the evolutionary terms $ M_*'$ and $ dlog(\phi)/dz$ with respect to the best-fit z=6 parameters of \citet{Mclure2009} (black point and errorbars). The
     upper and right axis refer to the  $ M_*$ and $ \phi$ values at the median redshift estimated for our sample (z=6.8). Grey points and errorbars mark the position of the $z\sim7$ best fit parameters by \citet{Ouchi2009} and by \citet{Bouwens2008}. The black solid line indicates the 99\% c.l. region estimated on the basis of the two GOODS-South pointings only (\C10).}
         \label{chi2}
\end{figure}

 We estimate how significant is the evolution of the LF at $z>6$ adopting a maximum likelihood approach. This method allows us to compare the observed number counts to those predicted for different evolving Schechter LFs  \citep{Schechter1976} after accounting for the expected systematics in
the detection process \cite[e.g.][]{Bouwens2007,Mannucci2007,Mclure2009}. As in \C10, we assume that the LF can be described by a Schechter
 function  with parameters $\phi$ and $M_*$ evolving from their value at $z_0=6.0$ \citep{Mclure2009} according to the following parametrisation:
$$log(\phi(z)) = log(\phi(z_0)) + dlog(\phi)/dz \cdot (z-z_0)$$
 $$M_*(z) = M_*(z_0) + M_*' \cdot (z-z_0)$$ Since our faint limit
 is close to the expected value of the characteristic luminosity
 $M_*$, we fix the faint end slope to the value $\alpha=-1.71$ of the $z \sim 6$ LF by \citet{Mclure2009}.
We  explicitly tested that no appreciable differences are found when fixing $\alpha$ to different values 
 ($\alpha = -1.4, -2.0$).
For a broad range of values of the evolving terms $ M_*'$ and $ dlog(\phi)/dz$ (see Fig.~\ref{chi2}) we
 simulate, under a MonteCarlo approach, the redshift $z$ and UV magnitude $M_{1500}$ for a population of $3\times10^5$ galaxies. These objects are randomly extracted from the larger database of simulated galaxies described in Sect.~ \ref{Montecarlo}, which encompasses a broad range of the physical parameters determining the rest frame photometry, like E(B-V), metallicity, Ly$\alpha$ EW etc.

The distributions of $Y$ magnitudes and $Z-Y$ colours for each simulated population are scaled to the observed area in each of the fields 
 and compared to the observed ones with a maximum
 likelihood test, under the assumption of simple Poissonian statistics.
For each of the two distributions, and for each field, we build the likelihood
 function $\cal{L}$:
 \begin{equation}
  {\cal L} = \prod_{i} e^{-N_{exp,i}} \frac{(N_{exp,i})^{N_{obs,i}}}{(N_{obs,i})!}
\label{eq:ml}
 \end{equation}
 where $N_{obs,i}$ is the observed number of sources in the magnitude
 (colour) interval $i$, $N_{exp,i}$ is the expected number of sources
 in the same magnitude (colour) interval, and $\Pi_{i}$ is the product
symbol. For each field, we associate to every model a likelihood computed as the
 product of those obtained for the magnitude and colour
 distributions separately. We then compute a final likelihood 
as the product of the GOODS, BDF and NTTDF likelihoods.

The colour plot in Fig.~ \ref{chi2} shows the 68\%, 95\%, and 99\% 
likelihood intervals on the evolutionary terms $ M_*'$ and $
dlog(\phi)/dz$ (left and bottom axes) and for the resulting Schechter
parameters at the median redshift z=6.8 of our sample (top and right
axes) for the combination of all the four Hawk-I fields.  In the same plot, the colour code refers to the $\chi^2$
distribution obtained under the usual assumption $\chi^2=-2.0\cdot
ln(\cal{L}) $ ~\cite[e.g.][]{Cash1979}.
\textit{We reject at $\gtrsim$ 99\% confidence level the hypothesis that the LF remains constant in both parameters above z=6 ($dlog(\phi)/dz=0$ and $M_*'=dM_*/dz=0$, black point in Fig.~ \ref{chi2})}. In Fig.~ \ref{chi2} it is also shown the 99\% c.l. region on the Schechter evolutionary terms estimated on the basis of the two GOODS-South pointings only (\C10). Although the degeneracy between $M_*$ and $\phi$ is still present, the analysis of the BDF and NTTDF fields considerably reduces the allowed parameter space.

The region of allowed values for the LF parameters in our final likelihood map points to a pronounced decrease of $\phi$ along with a brightening of $M_*$ with redshift.  However, the best-fit values for $M_*$ and $\phi$ at $z\sim7$ derived by \citet{Ouchi2009} and by \citet{Bouwens2008} \cite[see also][]{Bouwens2010c}, indicating a constant or slightly dimming $M_*$, still fall within the 2$\sigma$ region constrained by our maximum likelihood (grey points in  Fig.~ \ref{chi2}), and they are consistent with our estimate once the uncertainties are considered.   We argue that cosmic variance (see Sect.~\ref{cosmicvariance}) and the limited sample of very bright objects available may explain the discrepancies among different results: in our case, an inspection of the likelihood maps obtained separately on each field shows that the NTTDF, having two bright objects ($Y\sim25.5-25.7$, approximately $M_{1500}\lesssim-21.2$ at $z=6.8$), has a great effect in skewing the global likelihood towards brighter values of $M_*$. We also note that some theoretical models \cite[e.g.][]{Trenti2010,Finlator2010} predict a dimming of $M_*$ with redshift. However, several model parameters are largely unconstrained by the observations, while a large dust extinction might be required to match observed and predicted LFs at the bright end \citep{Lacey2010}.

\subsection{Cosmic Variance}
\label{cosmicvariance}
The effects of cosmic variance are reduced in our case,
since our data come from three independent areas, albeit of different sizes (the GOODS-South field being covered by two of the four Hawk-I pointings).
We evaluate the possible impact of cosmic variance using the mock catalogues of the Millennium Simulation \citep{kitzbichler2007} in the same way as discussed in \C10.
For each of the three Hawk-I areas (GOODS-South, BDF, NTTDF), we extract 200 fields of the same size from independent Millennium light-cones, and
we apply a corresponding photometric selection
criteria on galaxies at $6.5<z<7.4$ (bracketing the peak of our selection window), without any constraint on the
distribution of host haloes.
 We estimate that a cosmic
variance of $\sim 21\%$ affects the total number counts of $z$-drop LBGs in
our survey. We find that the evolution is still confirmed at a $\gtrsim$ 99\% confidence level by our maximum likelihood approach even allowing a $\sim 21\%$ variation in the total number density. Indeed, after accounting for all the observational effects, we estimate that we would have observed $\sim30$ ~$z$-drops in our survey in the case of a non-evolving LF: a factor of two higher than the observed number. However, while cosmic variance has not a significant effect on our conclusion that the LF strongly evolves from $z\sim7$ to $z\sim6$, it can have a great effect in determining the \textit{form} of this evolution. Cosmic variance is strongly luminosity dependent, and it is as high as 41\% for galaxies brighter than $Y=25.8$ in our survey, thus affecting the determination of the $M_*$ parameter.

\subsection{UV luminosity density, SFRD and constraints on cosmic reionization}
While the $M_*$ and $\phi$ parameters are highly degenerate, the number density of
 bright galaxies, i.e. the integral of the bright end of the LF is much better constrained, and so are derived integral quantities such as the  UV luminosity density ($\rho_{UV}$) and star formation rate density (SFRD).

We conservatively consider the model LFs within the 95\% c.l. region of our likelihood analysis to derive the $\rho_{UV}$ by integrating $L\cdot\Phi(L)$ up to the luminosity corresponding to $M_{1500}=-19.0$. 
We convert these values in a SFRD following
the standard formula by \citet{Madau1998} and applying the extinction
correction of \citet{Meurer1999} (considering an average UV slope
$\beta=-2.0$). Finally, we use  $\rho_{UV}$ to evaluate the emission rate  $\dot{N}_{ion}$ of hydrogen ionizing photons per $Mpc^3$  following \citet{Bolton2007}. We consider an escape fraction $f_{esc}=0.2$, a spectral index $\alpha_s=3.0$ and an ionizing emission density at the Lyman limit $\epsilon_g=\rho_{UV}/6.0$.

We report in Tab. ~\ref{Tab_parameters} the range of values for $\rho_{UV}$, SFRD and $log(\dot{N}_{ion})$.
These values are perfectly consistent with the analogous ones presented in \C10 and derived from the LFs in the 68\% c.l. region of the GOODS likelihood. Considering the same integral of the z=6 UV LF of \citet{Mclure2009}, our estimated $\rho_{UV}$ implies a drop of a factor
$\sim 3.5$ in the UV luminosity density from z=6. The \textit{lower limit} for the ionization rate required to balance recombination at $z=7$, computed according to \citet{Madau1999} and assuming an HII clumping factor equal to one, is $log(\dot{N}_{rec})=50.1$, which is a factor of two higher than the highest value allowed by our analysis. This demonstrates that, under usual assumptions, bright UV galaxies alone cannot keep the universe reionized at $z\sim7$.   By varying the escape  fraction, we obtain that values larger than $f_{esc}=0.5$ are required to reconcile the emission rate from bright galaxies with the one required for reionization. This demonstrates that, either UV bright galaxies at $z\sim7$ have different physical properties with respect to lower redshift LBGs, or, most probably, a crucial contribution to the reionization process comes from galaxies at the faint end of the LFs or from other kind of sources. Once different integration limits are taken into account, our estimates are in agreement with the results obtained by \citet{Bouwens2008,Ouchi2009,Gonzalez2010}.

\begin{table}
\caption{Properties of the $z\sim 7$ population$^a$}
\label{Tab_parameters}
\centering
\begin{tabular}{cc}
\hline

\hline

$\rho_{UV}$& $1.5 ^{+2.1} _{-0.8} ~10^{25}~erg ~s^{-1} ~Hz^{-1}~ Mpc^{-3}$\\
SFRD & $3.2 ^{+3.6} _{-1.9}~ 10^{-3}~ M_\odot ~yr^{-1}~ Mpc^{-3}$ \\
$log(\dot{N}_{ion})$& $49.4 ^{+0.4} _{-0.3} ~Mpc^{-3}$ \\
\hline
\end{tabular}
\\
\smallskip
\begin{tabular}{l}
a - LFs in the 95\% c.l. region ($M_{1500}<-19.0$) \\
\end{tabular}
\end{table}

\section{Constraints on the LBG number density at $z\sim8$}\label{z8}

We exploited deep Hawk-I $J$- and $K$-band observations to put an upper limit on the number density $z\gtrsim7.5$, $Y$-drop galaxies in our survey. We used the observations of the BDF and NTTDF fields presented in Sect.~\ref{dataset}, and deep observations of the two GOODS-South pointings obtained both in our program as well as through a similar ESO observing program (Cl\'ement et al. in preparation).  We obtained a multicolour catalogue with the $J$-band as detection image using SExtractor in dual mode over the full imaging set presented in this paper and in \C10. We used the same detection parameters and 2FWHM apertures adopted for the $Y$-detected catalogue, computing aperture corrected total magnitudes through appropriate corrections in each band. 
We chose the  colour selection criteria in order to isolate galaxies having the Lyman-break sampled by the $Y-J$ colour, and to exclude contamination from lower redshift galaxies on the basis of the expected colours for passive and dusty-starburst galaxies modelled as described in Sect.~\ref{ircolours}:

\begin{eqnarray*}
 (Y-J) &>& 0.8\\
(Y-J)&>& 1.1+0.6 \cdot (J-K)\\
\end{eqnarray*}

We also required no detection in the optical bands adopting the same $S/N$ criteria outlined for the selection of $z$-drop galaxies.
We limited our selection to $J$=24.5,24.8,25.0 in the BDF, NTTDF and GOODS pointings respectively, up to which we estimate that our catalogues are 100\% complete, and we used the same area chosen for the selection of $z$-drop galaxies in order to avoid the noisiest regions in any image.

With these criteria we found no candidate $Y$-drop galaxy in our survey. 
Considered the average $J-M_{1500}$ relation at the median redshift of our colour selection (z=8), we are probing the $M < M_*$ region of the LF at $M_{1500}\sim -22.5$. We  report in Tab.~\ref{Tab_z8}  an upper limit on the number density of very bright $Y$-drop LBGs estimated as the inverse of the volume sampled by our survey in the redshift interval $7.5<z<9.0$.

\begin{table}
\caption{LBG number density at $z\sim8$}
\label{Tab_z8}
\centering
\begin{tabular}{cc}
\hline
Mag. Range & $\phi ~ (10^{-4} Mpc^{-3})$\\
\hline
$M_{1500}<-22.0$& $<0.02$\\
\hline
\end{tabular}
\end{table}
 
\section{Summary and conclusions}

We presented in this work the results of a $Y$--band survey of the
two high galactic latitude BDF and NTTDF fields aimed at detecting galaxies at $z \gtrsim 6.5$ and
measuring their number density. The survey is based on deep observations obtained under a dedicated ESO Large Programme. We made use of $Y$, $J$, $K$ band observations
performed with Hawk-I, the new near-IR camera
installed at the VLT, and of FORS2 $Z$-band observations. We matched and combined these data with deep archive FORS1 and FORS2 observations in the $U$, $B$, $V$, $R$, $I$ filters to detect high redshift LBGs under the main criterion $Z-Y>1$, requiring no optical detection and flat $Y-J$ and $Y-K$ colours.
The  colour selection criteria have been tailored in order to exclude
lower redshift passive galaxies and dusty starbursts, Galactic T-dwarfs and galaxies exhibiting large $Z-Y$
 colours as well as significant emission in the optical bands, possibly intermediate redshift sources with bright emission lines.

As a result, we isolated 8 highly reliable $z$-drop candidates in the magnitude range $Y\simeq 25.5-26.5$ over a total area of 70.1 $arcmin^2$. 
We combined this $z$-drop sample with the similar one extracted from two pointings over the GOODS-South field comprising seven galaxies at $Y<26.7$. 

We performed a stacking analysis of the 15 objects to estimate the average properties of $M\sim M_*$ galaxies at $z\gtrsim 6.5$. The photometric
redshift of the stacked object is $z=6.85^{+0.20}_{-0.15}$ in perfect agreement with the estimated selection window of our survey. The stacked SED is fitted by an $E(B-V)=0.05^{+0.15}_{-0.05}$ at a 68\% confidence level, indicating a low dust content in agreement with previous analysis of z$\sim7-8$  objects \citep{Finkelstein2009,Schaerer2010,Gonzalez2010}.

We then estimated the number density and the LF evolution on the basis of detailed MonteCarlo imaging simulations accounting for
 all the uncertainties involved in the observations: detection
 completeness, photometric scatter, and random fluctuations in the S/N
 measure due to overlapping unresolved sources, or other effects. 
We first computed a binned estimate of the galaxy number density at $z\sim7$ following two different procedures. The first one, which is based on an average $Y$-$M_{1500}$ relation and on an estimate of the redshift dependent completeness of our selection, is the procedure commonly adopted in the literature. The second method is more conservative, and exploits a $\chi^2$ minimization to compare the observed  number counts to those predicted on the basis of MonteCarlo simulations for different combinations of galaxy densities. This second procedure intrinsically considers the uncertainties in the $Y$-$M_{1500}$ conversion due to photometric scatter, to the redshift distribution and to the intrinsic properties of different galaxy models. We find that the two procedures are consistent and they are in agreement with similar analysis from the literature. However, the more conservative procedure highlights that sources of statistical uncertainty are usually underestimated.

To assess the degree of evolution
 of the UV LF at $z>6.0$, we also simulated galaxy populations following different UV
 Schechter functions with linearly evolving parameters $log(\phi)$ and
 $M_*$.  For each of the four Hawk-I pointings we
 compared the resulting distributions of simulated magnitudes and
 colours with the observed ones following a maximum likelihood approach.
 We find strong evidence of evolution of the LF
 above z=6: our analysis rules out at a $> 99\%$ confidence level
 that the LF remains constant in both $\phi$ and $M_*$ above $z=6$. 
Our likelihood maps for the Schechter parameters indicate a strong evolution in $\phi$ and a brightening of $M_*$ with redshift. However, the detection of two bright objects ($Y\sim 25.5-25.7$, corresponding to $M_{1500}\lesssim-21.2$) in the NTTDF pointing have a major role in skewing the evolution of $M_*$ towards bright values. The two Schechter parameters are, however, highly degenerate and our findings are also consistent within the uncertainties with a milder evolution of $\phi$ and a constant or slightly dimming $M_*$ as indicated by other authors \citep{Bouwens2008,Ouchi2009}. We estimate that the possible effect of cosmic variance is not capable of reconciling the observed number density of $z$-drop galaxies with the one predicted for a non-evolving LF. However, the strong dependence on luminosity of the cosmic variance, and the relatively small magnitude range probed by our survey at $M\lesssim M_*$, can influence the determination of the form of the evolving LF and provide an explanation for the difference between the evolution we determine and other estimates in the literature.

 The uncertainty and the degeneracy in the $M_*$ and
 $\phi$ best-fit values are not reflected in a comparable uncertainty
 in the number density of bright galaxies.  We conservatively consider the model LFs within the 95\% c.l. region of our likelihood analysis to derive for galaxies at $M_{1500}<-19.0$  an UV luminosity density $\rho_{UV}=
 1.5^{+2.1}_{-0.8} 10^{25} erg ~ s^{-1} ~ Hz^{-1} ~ Mpc^{-3} $, a star formation rate density $SFRD=3.2 ^{+3.6} _{-1.9}~
 10^{-3} M_{\odot} ~yr^{-1} ~ Mpc^{-3} $ and an emission rate of hydrogen ionizing photons $log(\dot{N}_{ion})=49.4 ^{+0.4} _{-0.3} ~Mpc^{-3}$. The UV luminosity density is lower than the corresponding one at $z\sim 6$ by a factor $\sim 3.5$,  while $\dot{N}_{ion}$ is lower by at least a factor of  $\sim 2$ than the lower limit required for reionization according to \citet{Madau1999}, considering $f_{esc}=0.2$ and an HII clumping factor equal to one. This implies that UV bright galaxies alone cannot reionize the universe, unless their physical parameters are much different from those of lower redshift LBGs (e.g. $f_{esc}>0.5$, harder UV spectrum etc.). Most probably, the crucial contribution to reionization comes from galaxies at the faint end of the LF or from other kind of sources. 
Finally, we exploit the Hawk-I $J$ and $K$ band observations of our survey to derive an upper limit of $2\cdot10^{-6} Mpc^{-3}$ for the number density of  $M \sim -22.5$ LBGs at $z\sim8$ from the non-detection of $Y$-drop galaxies up to $J\sim25$.

\begin{acknowledgements}

  Observations were carried out using the Very Large Telescope at
  the ESO Paranal Observatory under Programme IDs LP181.A-0717,
  LP168.A-0485, ID 170.A-0788, ID 181.A-0485, ID 283.A-5052 and the ESO Science Archive under
  Programme IDs 67.A-0249, 71.A-0584, 73.A-0564, 68.A-0563, 69.A-0539, 70.A-0048, 64.O-0643, 66.A-0572, 68.A-0544, 164.O-0561, 163.N-0210,
  and 60.A-9120. We acknowledge support from Agenzia Spaziale Italiana.
\end{acknowledgements}

\bibliographystyle{aa}

\end{document}